\documentclass[aps,pra,onecolumn,groupedaddress,showpacs]{revtex4}

\usepackage{amssymb,amsmath}
\usepackage{graphicx}

\newcommand{\om}{\omega}

\newcommand{\pa}{\partial}

\begin{document}

\title{Two-component nonlinear wave of the NLS equation}

\author{G. T. Adamashvili}
\affiliation{Technical University of Georgia, Kostava str.77, Tbilisi, 0179, Georgia.\\ email: $guram_{-}adamashvili@ymail.com.$ }

\begin{abstract}
Using the generalized perturbation reduction method the scalar nonlinear Schr\"odinger equation is transformed to the coupled nonlinear Schr\"odinger equations for auxiliary functions. A solution in the form of a two-component vector nonlinear pulse is obtained. The components of the pulse oscillate with the sum and difference of the frequencies and wave numbers.
Explicit analytical expressions for the shape and parameters of the two-component nonlinear pulse are presented.
$$$$
\end{abstract}


\maketitle

\section{Introduction}

The  nonlinear solitary wave solutions (soliton, breather, vector breather and etc.) of various nonlinear partial differential equations  plays a fundamental role in the theory of nonlinear waves and applied mathematics. These equations include the Sine-Gordon equation, the Maxwell-Bloch equations, different versions of the modified Benjamin-Bona-Mahony equations, the Maxwell-Liouville equations,  the system of the magnetic Bloch equations and the acoustic wave equation, the wave equation in dispersive and Kerr-type media, the system of wave equation and material equations for multi-photon resonant excitations, among others. To obtain the solitary wave solutions to these nonlinear evolution equations, many methods were developed, such as the inverse scattering transform, different asymptotical approaches, Lie group method, the factorization technique, Exp-function method and so on [1-9].
Special interest express the single (scalar) and coupled nonlinear Schr\"odinger equations. They are being one of the basic equations for studying the one-component and two-component solitary waves of stable profile of any nature in different physical fields of research: in optics, acoustics, magnetics, fluid dynamics, quantum electronics, particle physics, plasma physics and etc. By means of various mathematical approaches we can establish a connection between two different nonlinear equations. In particular, to apply the generalized perturbation reduction method we can transform various nonlinear partial differential equations to the coupled nonlinear Schr\"odinger equations for auxiliary functions. As a result, the two-component waves, which are the bound state of two small-amplitude scalar one-component breathers with the identical polarization, have been obtained. The first breather oscillates with the sum, and the second with the difference of the frequencies and wave numbers. This two-component pulse has a very special form and they are met in different fields of research: in optics, acoustics, plasma physics, hydrodynamics, particle physics, and etc. In the optical and acoustic self-induced transparency, such wave is called the vector $0\pi$ pulse [10-24].

It is necessary to separately consideration the nonlinear partial differential equations for real and complex functions.
We consider the real function $U(z,t)$ of the space coordinate $z$ and time $t$  which can be presented to the form
\begin{equation}\label{u}
 U=U^{(+)}+U^{(-)},
\end{equation}
where the complex conjugate functions $U^{(+)}$  and $U^{(-)}$ are satisfied the scalar nonlinear Schr\"odinger equation
\begin{equation}\label{nse 1}
\pm i \frac{\partial U^{(\pm)}}{\partial t} +\beta\frac{\partial^{2} U^{(\pm)}}{\partial z^2}+a |U^{(\pm)}|^{2} U^{(\pm)}=0,
\end{equation}
$U^{(+)}$=${U^{(-)}}^{*}$, $a$ and $\beta$ are real constants. Eq.(2) is completely integrable by the inverse scattering transform and has $N$-soliton solutions [1-4].

One-soliton solution of this equation has the form
\begin{equation}\label{soli}
U^{(\pm)}(z,t)= K\frac{e^{\pm i  \phi_{1}(z,t)}} {cosh\phi_{2}(z,t)},
\end{equation}
where
\begin{equation}\label{phi}\nonumber
\phi_{1} (z,t)=\frac{1}{2}[\frac{V_s}{\beta}z-(\frac{V^{2}_{s}}{2\beta}-a K^{2})t],
$$
$$
\phi_{2} (z,t)={K\sqrt{\frac{a}{2\beta}}(z-V_{s}t)},
\end{equation}
$K$  is the amplitude  of the nonlinear Schr\"odinger equation soliton, $V_s$ is the velocity of the nonlinear wave.

Depending from the width of the nonlinear pulse we can consider different nonlinear solitary wave solutions.
In the present work we investigate one-component scalar breather and two-component vector breather solutions of Eqs.(1) and (2).

The rest of this paper is organized as follows: Section II is devoted to the small amplitude one-component breather solution.
In Section III, we will investigate the two-component (vector) breather solution of the Eqs.(1) and (2) with the generalized perturbation reduction method.
Finally, in Section IV, we will discuss the obtained results.

\vskip+0.5cm
\section{The one-component breather solution}

When the width of the pulse $T>>\omega^{-1}$ for the real function $U(z,t)$  Eq.(1) we can using the method of slowly varying shape.
In this case we represent the functions $U^{(\pm)}$ in the form
\begin{equation}\label{upm}
U^{(\pm)}= \hat{u}_{\pm 1}Z_{\pm 1},
\end{equation}
where $\hat{u}_{\pm 1}$ is the slowly varying complex  envelope function, $Z_{\pm 1}= e^{\pm i(kz -\om t)}$ is the fast oscillating function, $\omega$ and $k$ are the frequency and the wave number of the carrier wave.  For the reality of $U$, we set $ \hat{u}_{+1}= \hat{u}^{*}_{-1}$.

The complex function $\hat{u}_{\pm 1}$ vary sufficiently slowly in space and time compared to the carrier wave part $Z_{\pm 1}$, i.e.
\begin{equation}\nonumber
|\frac{\partial \hat{u}_{\pm 1}}{\partial t}|<<\omega |\hat{u}_{\pm 1}|,\;\;\;\;\;\;\;\;\;\;\;\;\; |\frac{\partial \hat{u}_{\pm 1}}{\partial z}|<< k |\hat{u}_{\pm 1}|
\end{equation}
are valid.

Substituting Eq.(4) into Eq.(2), we obtain the  nonlinear Schr\"odinger equation for  envelope function $\hat{u}_{\pm 1}$ in the form
\begin{equation}\label{eqp}
\pm i (\frac{\pa \hat{u}_{\pm 1}}{\pa t}+ 2 \beta k \frac{\pa \hat{u}_{\pm1} }{\pa z}) +\beta \frac{\pa^{2} \hat{u}_{\pm1}}{\pa z^2}=-a |\hat{u}_{\pm1}|^{2} \hat{u}_{\pm1}
\end{equation}
and dispersion relation for nonlinear pulse
\begin{equation}\label{dis1}
 \om =\beta k^{2}.
\end{equation}

The equation (5) is transformed to the scalar nonlinear Schr\"odinger equation in the following form
 \begin{equation}\label{nse5}
\pm i \frac{\pa \hat{u}_{\pm 1}}{\pa t} +\beta \frac{\pa^{2} \hat{u}_{\pm1}}{\pa y^2}  +a |\hat{u}_{\pm1}|^{2} \hat{u}_{\pm1}=0,
\end{equation}
where
$$
y=z-2 \beta k \;t,\;\;\;\;\;\;\;\;\;\;t=t.
$$

The one-soliton solution of the Eq.(7) is given by the Eq.(3) if in this equation replace $z$ by  the variable $y$.
Substituting this solution to the Eqs.(1) and (4) we obtain
\begin{equation}\label{b}
U= 2 K\frac{\sin(k z-\omega t+\phi_{1}(y,t)-\frac{\pi}{2})}{\cosh \phi _{2}(y,t)}.
\end{equation}
This is small amplitude single-component breather for the function $U$.

We have to note that Eq.(8) is not breather of the nonlinear Schr\"odinger equation (7). The breather of nonlinear Schr\"odinger equation is unstable solution to relatively infinitesimal perturbations of the initial data. The real function $U$ can be solution of the another nonlinear equation, for instance, the Sin-Gordon equation. In this case, we obtain well known result when one soliton solution of nonlinear Schr\"odinger equation is connected with the small amplitude breather solution of the Sine-Gordon equation[1].

\vskip+0.5cm
\section{The two-component vector breather solution}

The single-component breather Eq.(8) is not the only possible nonlinear wave for the function
\begin{equation}\label{ub}
 U=\sum_{l=\pm1}\hat{u}_{l}Z_{l},
\end{equation}
when envelope of this function $\hat{u}_{l}$ satisfied scalar nonlinear Schr\"odinger equation Eq.(5).

Indeed, for more wider pulses for which  the condition
\begin{equation}\label{oO}
\omega >> \Omega_{l,n} >> T^{-1}
\end{equation}
is fulfilled, we can  consider  also the two-component nonlinear solitary waves, where parameters  $\Omega_{l,n}$ will be determined below.

For the study of the two-component nonlinear solitary wave solution of Eqs.(5) and (9) we apply the generalized perturbative reduction method [10-18] by means of which we can transform  Eq.(5) into the coupled nonlinear Schr\"odinger equations for auxiliary functions. In this method the function $\hat{u}_{l}(z,t)$ can be represented as:
\begin{equation}\label{cemi}
\hat{u}_{l}(z,t)=\sum_{\alpha=1}^{\infty}\sum_{n=-\infty}^{+\infty}\varepsilon^\alpha
Y_{l,n} f_{l,n}^ {(\alpha)}(\zeta_{l,n},\tau),
\end{equation}
where
$$
Y_{l,n}=e^{in(Q_{l,n}z-\Omega_{l,n} t)},\;\;\;\zeta_{l,n}=\varepsilon Q_{l,n}(z-{v_g}_{l,n} t),
$$$$
\tau=\varepsilon^2 t,\;\;\; \;\;\;\;\;\;\;\;\;\;\;{ v_{g;}}_{l,n}=\frac{d\Omega_{l,n}}{dQ_{l,n}},
$$
$\varepsilon$ is a small parameter. Such an expansion allows us to separate from $\hat{u}_{l}$ the even more slowly changing auxiliary function $ f_{l,n}^{(\alpha )}$. It is assumed that the quantities $\Omega_{l,n}$, $Q_{l,n}$, and $f_{l,n}^{(\alpha)}$ satisfy the conditions:
\begin{equation}\label{rpt}\nonumber
\omega\gg \Omega_{l,n},\;\;k\gg Q_{l,n},
\end{equation}
$$
\left|\frac{\partial
f_{l,n}^{(\alpha )}}{
\partial t}\right|\ll \Omega_{l,n} \left|f_{l,n}^{(\alpha )}\right|,\;\;\left|\frac{\partial
f_{l,n}^{(\alpha )}}{\partial z }\right|\ll Q_{l,n}\left|f_{l,n}^{(\alpha )}\right|.
$$
for any value of indexes $l$ and $n$.

Substituting Eq.(11) into the left-hand side of the Eq.(5), we obtain wave equation
\begin{equation}\label{eqw}
 \sum_{\alpha=1}^{\infty}\sum_{n=-\infty}^{+\infty}\varepsilon^\alpha
Y_{l,n} [W_{l,n} +i\varepsilon J_{l,n} \frac{\partial }{\partial \zeta}
+il\varepsilon^2 \frac{\partial }{\partial \tau}
+ \beta \varepsilon^{2} Q^{2} \frac{\partial^2 }{\partial \zeta^2}]f_{l,n}^{(\alpha)} +a |\hat{u}_{l}|^{2} \hat{u}_{l}=0,
\end{equation}
where
\begin{equation}\nonumber
W_{l,n}=l n\Omega_{l,n} -2l \beta k n Q_{l,n} -n^2\beta   Q^{2}_{l,n},
$$$$
J_{l,n}=  -l Q_{l,n}  { v_{g;}}_{l,n} + 2 l \beta k  Q_{l,n} + 2\beta  n Q^{2}_{l,n}.
\end{equation}

Following the standard procedure characterized for any perturbative expansions while equating to each other the
terms of the same order to $\varepsilon$, from the Eq.(12), we obtain the chain of the equations. As a
result, in the first and second order of $\varepsilon$ we determine the connection between the parameters $\Omega_{l,n}$ and $Q_{l,n}$ which is given by
\begin{equation}\label{dis22}
 \Omega_{l,n} = \beta Q_{l,n} (2 k +nl Q_{l,n})
\end{equation}
and the equation
\begin{equation}\label{vv}
{ v_{g;}}_{l,n}=2\beta (k +  nl Q_{l,n}).
\end{equation}

In the third order of $\varepsilon$ we obtain the coupled nonlinear Schr\"odinger equations  in the following form
\begin{equation}\label{cnse}
i (\frac{\partial \lambda_{\pm 1}}{\partial t}+ v_{\pm}\frac{\partial  \lambda_{\pm 1}} {\partial z})+\beta  \frac{\partial^{2} \lambda_{\pm 1} }{\partial z^{2}} +  a    (  |\lambda_{\pm 1}|^{2} + 2 |\lambda_{\mp 1}|^{2} )\lambda_{\pm 1}=0,
\end{equation}
where
\begin{equation}\label{labda}\nonumber
\lambda_{\pm 1}=\varepsilon  f_{+1,\pm 1}^{(1)},
$$$$
  v_{\pm}= { v_{g;}}_{+1,\pm1}= 2  \beta (k  \pm Q_{+1,\pm1}).
\end{equation}

The solution of Eq.(15) we seek in the form of [10-18] 
\begin{equation}\label{ue1}
\lambda_{\pm 1}=\frac{K_{\pm }}{\mathfrak{b} T}Sech(\frac{t-\frac{z}{V_{0}}}{T}) e^{i(k_{\pm } z - \omega_{\pm } t )},
\end{equation}
where $K_{\pm },\; k_{\pm }$ and $\omega_{\pm }$ are the real constants, $V_{0}$ is the velocity of the nonlinear wave. We assume that
\begin{equation}\label{kom}
k_{\pm }<<Q_{+1,\pm 1},\;\;\;\;\;\;\omega_{\pm }<<\Omega_{+1,\pm 1}.
\end{equation}

The connections between  $K_{\pm },\; k_{\pm }$ and $\omega_{\pm }$  are given by
\begin{equation}\label{ttw}
K_{+}^{2}=K_{-}^{2},\;\;\;\;\;\;\;\;\;\;\;\;\;\;\;k_{\pm }=\frac{V_{0}-v_{\pm}}{2\beta},
$$$$
\omega_{+}=\omega_{-}+\frac{v_{-}^{2}-v_{+}^{2}}{4\beta}.
\end{equation}

Substituting Eq.(16) into (15), when the complex envelope functions $\hat{u}_{\pm 1}$ satisfied the scalar nonlinear Schr\"odinger equation (7), we obtain the two-component pulse of the function $U$:
\begin{equation}\label{vb}
U(z,t)=\frac{2 }{b T}Sech(\frac{t-\frac{z}{V}}{T})\{  K_{+} \cos[(k+Q_{+1,+1}+k_{+})z
-(\omega +\Omega_{+1,+1}+\omega_{+}) t] $$$$ +K_{-}\cos[(k-Q_{+1,-1}+k_{-})z -(\omega -\Omega_{+1,-1}+\omega_{-})t]\},
\end{equation}
where $T$ is the width of the two-component nonlinear pulse,
\begin{equation}\label{rrw}
T^{-2}=V_{0}^{2}\frac{v_{+}k_{+}+k_{+}^{2}\beta -\omega_{+}}{\beta},
$$$$
\mathfrak{b}^{2}=\frac{3 V_{0}^{2}a}{2\beta}K_{+}^{2} .
\end{equation}
This pulse oscillating with the sum $\omega+\Omega_{+1,+1}$
and difference $\omega-\Omega_{+1,-1}$ of the frequencies and the wave numbers $k+Q_{+1,+1}$ and $k-Q_{+1,-1}$ (taking into account Eq.(17)).

\vskip+0.5cm
\section{Conclusion}

In the present paper, we investigate the wide class of phenomena for the nonlinear waves of different nature (optical, acoustic, elastic, magnetic and etc.) in various  media and different fields of research which can be describe by means of the scalar nonlinear Schr\"odinger equation. We study separately nonlinear waves with different width and is shown that can be formed different nonlinear waves: soliton Eq.(3), and for more wider pulse small amplitude breather Eq.(8), which coincide with the small amplitude breather of Sine-Gordon equation.

For more wider width of pulses for which  the inequalities (10) is valid, situation became different. Using the generalized perturbation reduction method Eq.(11), the
Eq.(5) for the complex functions $\hat{u}_{l} $,  transform to the coupled nonlinear Schr\"odinger equations (15) for the auxiliary functions $\lambda_{\pm 1}$. As a result, the two-component nonlinear pulse oscillating with the sum and difference  of the frequencies and wave numbers Eq.(19), can be formed. The dispersion relation and the connection between parameters $Q_{+1,\pm 1}$ and $\Omega_{+1,\pm 1}$ are determined from Eqs.(6) and (13).
The parameters of the two-component vector breather of Eqs.(1) and (2) are determined from Eqs.(14), (18) and (20).

\vskip+0.5cm


\begin{thebibliography}{20}
\expandafter\ifx\csname natexlab\endcsname\relax\def\natexlab#1{#1}\fi
\expandafter\ifx\csname bibnamefont\endcsname\relax
  \def\bibnamefont#1{#1}\fi
\expandafter\ifx\csname bibfnamefont\endcsname\relax
  \def\bibfnamefont#1{#1}\fi
\expandafter\ifx\csname citenamefont\endcsname\relax
  \def\citenamefont#1{#1}\fi
\expandafter\ifx\csname url\endcsname\relax
  \def\url#1{\texttt{#1}}\fi
\expandafter\ifx\csname urlprefix\endcsname\relax\def\urlprefix{URL }\fi
\providecommand{\bibinfo}[2]{#2}
\providecommand{\eprint}[2][]{\url{#2}}



%(1)
\bibitem[{\citenamefont{Newell}(1975)}]{Newell::85}
\bibinfo{author}{\bibfnamefont{A.~C.}~\bibnamefont{Newell}},
\emph{\bibinfo{title}{\emph{Solitons in Mathematics and Physics }}}
(\bibinfo{publisher}{Society for Industrial and Applied Mathematics}, \bibinfo{year}{1985}).

%(2)
\bibitem[{\citenamefont{Novikov et~al.}(1984)\citenamefont{Novikov}}]{Novikov::84}
\bibinfo{author}{\bibfnamefont{S.~P.} \bibnamefont{Novikov}},
\bibinfo{author}{\bibfnamefont{S.~V.} \bibnamefont{Manakov}}
\bibinfo{author}{\bibfnamefont{L.~P.} \bibnamefont{Pitaevski}} \bibnamefont{and}
\bibinfo{author}{\bibfnamefont{V.~E.} \bibnamefont{Zakharov}} ,
\bibinfo{journal}{\emph{Theory of Solitons: The Inverse Scattering Method}, (Academy of Science of the USSR, Moscow, USSR. 1984).}

%(3)
\bibitem[{\citenamefont{Dodd}(2006)}]{Dodd::1982}
\bibinfo{author}{\bibfnamefont{R.~K.} \bibnamefont{Dodd}},
\bibinfo{author}{\bibfnamefont{J.~C.} \bibnamefont{Eilbeck}},
\bibinfo{author}{\bibfnamefont{J.~D.} \bibnamefont{Gibbon}} \bibnamefont{and}
\bibinfo{author}{\bibfnamefont{H.~C.} \bibnamefont{Morris}}
\bibinfo{journal}{\emph{Solitons and Nonlinear wave Equations}, Academic Press. Inc.} (\bibinfo{year}{1982}).

%(4)
\bibitem[{\citenamefont{Ablowitz}(1974)}]{Ablowitz::81}
\bibinfo{author}{\bibfnamefont{M.~J.}\bibnamefont{Ablowitz}} \bibnamefont{and}
\bibinfo{author}{\bibfnamefont{H.} \bibnamefont{Segur}},
\bibinfo{journal}{\emph{Solitons and Inverse Scattering Transform}},
\bibinfo{pages}{(SIAM Philadelphia)} (\bibinfo{year}{1981}).

%(5)
\bibitem[{\citenamefont{Sauter E.G.}(1996}]{Sauter::96}
\bibinfo{author}{\bibfnamefont{E.~G.} \bibnamefont{Sauter}},
  \emph{\bibinfo{title}{Nonlinear Optics}}
  (\bibinfo{publisher}{Wiley, New york,} \bibinfo{year}{1996}).

%(6)
\bibitem[{\citenamefont{ Taniuti}(1973)}]{Taniuti::1973}
\bibinfo{author}{\bibfnamefont{T.} \bibnamefont{Taniuti}} \bibnamefont{and}
\bibinfo{author}{\bibfnamefont{N.} \bibnamefont{Iajima}},
\bibinfo{journal}{J. Math. Phys.} \textbf{\bibinfo{volume}{14}},
\bibinfo{pages}{1389} (\bibinfo{year}{1973}).

%(7)
\bibitem[{\citenamefont{ Leblond}(2008)}]{Leblond::08}
\bibinfo{author}{\bibfnamefont{H.}~ \bibnamefont{Leblond}},
\bibinfo{journal}{J.Phys.B.} \textbf{\bibinfo{volume}{41}},
\bibinfo{pages}{043001} (\bibinfo{year}{2008}).

%(8)
\bibitem[{\citenamefont{Witham }(1984)}]{Witham::74}
\bibinfo{author}{\bibfnamefont{G.~G.}~\bibnamefont{Witham }}
\emph{\bibinfo{title}{Linear and Nonlinear Waves.}}
(\bibinfo{publisher}{John Wiley, New York.}, \bibinfo{year}{1974}).

%(9)
\bibitem[{\citenamefont{ Singh}(1973)}]{Singh::06}
\bibinfo{author}{\bibfnamefont{K.} \bibnamefont{Singh}}
\bibnamefont{and}
\bibinfo{author}{\bibfnamefont{R.~K.} \bibnamefont{Gupta}},
\bibinfo{journal}{International Journal of Engineering Science, } \textbf{\bibinfo{volume}{44}},
\bibinfo{pages}{241} (\bibinfo{year}{2006}).

%(10)
\bibitem[{\citenamefont{Adamashvili }(2011)\citenamefont{Adamashvili}}]{Adamashvili:Result:11}
\bibinfo{author}{\bibfnamefont{G.~T.} \bibnamefont{Adamashvili}},
    \bibinfo{journal}{Results in  Physics},  \textbf{\bibinfo{volume}{1}},
  \bibinfo{pages}{26} (\bibinfo{year}{2011}).

%(11)
\bibitem[{\citenamefont{Adamashvili}(2012)\citenamefont{Adamashvili }}]{Adamashvili:Optics and spectroscopy:2012}
\bibinfo{author}{\bibfnamefont{G.~T.} \bibnamefont{Adamashvili}},
\bibinfo{journal}{Optics and spectroscopy,}  \textbf{\bibinfo{volume}{113}},
\bibinfo{pages}{1} (\bibinfo{year}{2012}).

%(12)
\bibitem[{\citenamefont{Adamashvili et~al.}(2012)}]{Adamashvili:Physica B:12}
\bibinfo{author}{\bibfnamefont{G.~T.} \bibnamefont{Adamashvili}},
\bibinfo{journal}{Physica B.}  \textbf{\bibinfo{volume}{407}},
\bibinfo{pages}{3413} (\bibinfo{year}{2012}).

%(13)
\bibitem[{\citenamefont{Adamashvili et~al.}(2012)}]{Adamashvili:Eur.Phys.J.D.:12}
\bibinfo{author}{\bibfnamefont{G.~T.} \bibnamefont{Adamashvili}},
\bibinfo{journal}{The Eur. Phys. J. D.}  \textbf{\bibinfo{volume}{66}},
\bibinfo{pages}{101} (\bibinfo{year}{2012}).

%(14)
 \bibitem[{\citenamefont{Adamashvili }(2016)\citenamefont{Adamshvili }}]{Adamshvili:Arxiv:2014}
\bibinfo{author}{\bibfnamefont{G.~T.} \bibnamefont{Adamashvili}},
\bibinfo{author}{\bibfnamefont{M.~D.} \bibnamefont{Peikrishvili}},
\bibinfo{author}{\bibfnamefont{R.~R.} \bibnamefont{Koplatadze}} \bibnamefont{and}
\bibinfo{author}{\bibfnamefont{K.~L.} \bibnamefont{Schengelia}},
    \bibinfo{journal}{ Arxiv: 1408.4310v1,}
 (\bibinfo{year}{19 Aug 2014}).

%(15)
\bibitem[{\citenamefont{Adamashvili }(2015)}]{Adamashvili:PhysLettA:2015}
\bibinfo{author}{\bibfnamefont{G.~T.} \bibnamefont{Adamashvili}},
  \bibinfo{journal}{Phys. Lett. A} \textbf{\bibinfo{volume}{379}},
\bibinfo{pages}{218}  (\bibinfo{year}{2015}).

%(16)
\bibitem[{\citenamefont{Adamashvili}(2019)\citenamefont{Adamashvili }}]{Adamashvili:Optics and spectroscopy:2019}
\bibinfo{author}{\bibfnamefont{G.~T.} \bibnamefont{Adamashvili}},
\bibinfo{journal}{Optics and spectroscopy,}  \textbf{\bibinfo{volume}{127}},
\bibinfo{pages}{865} (\bibinfo{year}{2019}).

%(17)
\bibitem[{\citenamefont{Adamashvili }(2019)\citenamefont{Adamshvili,  }}]{Adamshvili:Arxiv:2019}
\bibinfo{author}{\bibfnamefont{G.~T.} \bibnamefont{Adamashvili}},
    \bibinfo{journal}{Arxiv: 1907.10883v1,}
 (\bibinfo{year}{25 Jul 2019}).

%(18)
\bibitem[{\citenamefont{Adamashvili et~al.}(2020)\citenamefont{Adamashvili,
Weber, Knorr, and Adamashvili}}]{Adamashvili:Eur.Phys.J.D.:20}
\bibinfo{author}{\bibfnamefont{G.~T.} \bibnamefont{Adamashvili}},
\bibinfo{journal}{The Eur. Phys. J. D},  \textbf{\bibinfo{volume}{74}},
\bibinfo{pages}{Issue 3, 41} (\bibinfo{year}{2020}).

%(19)
\bibitem[{\citenamefont{Adamashvili }(2012)}]{Adamashvili:PRE:12}
\bibinfo{author}{\bibfnamefont{G.~T.} \bibnamefont{Adamashvili}},
  \bibinfo{journal}{Phys. Rev. E},  \textbf{\bibinfo{volume}{85}},
  \bibinfo{pages}{067601} (\bibinfo{year}{2012}).

%(20)
\bibitem[{\citenamefont{Adamashvili}(2016)\citenamefont{Adamashvili }}]{Adamashvili:PRE:16}
\bibinfo{author}{\bibfnamefont{G.~T.} \bibnamefont{Adamashvili}},
   \bibinfo{journal}{Phys. Rev. E}  \textbf{\bibinfo{volume}{93}},
  \bibinfo{pages}{023002} (\bibinfo{year}{2016}).

%(21)
 \bibitem[{\citenamefont{Adamashvili }(2020)\citenamefont{Adamshvili,  }}]{Adamshvili:Arxiv:2020}
\bibinfo{author}{\bibfnamefont{G.~T.} \bibnamefont{Adamashvili}},
    \bibinfo{journal}{Arxiv: 2001.07758v1,}
 (\bibinfo{year}{21 Jan 2020}).

%(22)
\bibitem[{\citenamefont{Adamashvili }(2019)\citenamefont{Adamshvili,  }}]{Adamshvili:Arxiv:2018}
\bibinfo{author}{\bibfnamefont{G.~T.} \bibnamefont{Adamashvili}},
\bibinfo{author}{\bibfnamefont{N.~T.} \bibnamefont{Adamashvili}},
\bibinfo{author}{\bibfnamefont{M.~D.} \bibnamefont{Peikrishvili}} \bibnamefont{and}
\bibinfo{author}{\bibfnamefont{R.~R.} \bibnamefont{Koplatadze}},
    \bibinfo{journal}{Arxiv:1804.02993v1,}
 (\bibinfo{year}{6 Apr 2018}).

%(23)
\bibitem[{\citenamefont{Adamashvili}(2017)\citenamefont{Adamashvili }}]{Adamashvili:AcousPhy:17}
\bibinfo{author}{\bibfnamefont{G.~T.} \bibnamefont{Adamashvili}},
   \bibinfo{journal}{Acoustical Physics,}  \textbf{\bibinfo{volume}{63}},
  \bibinfo{pages}{517} (\bibinfo{year}{2017}).

%(24)
 \bibitem[{\citenamefont{Adamashvili }(2016)\citenamefont{Adamshvili,  }}]{Adamshvili:Arxiv:2016}
\bibinfo{author}{\bibfnamefont{G.~T.} \bibnamefont{Adamashvili}},
    \bibinfo{journal}{Preprint, Arxiv: 1608.05349v1,}
 (\bibinfo{year}{18 Aug 2016}).








































\end{thebibliography}
\end{document}